\documentclass[11pt]{article}
\begin{document}
  
\title{{\bf New Nonlocal Charges in SUSY Integrable Models}}
\author{Ashok Das\\
Department of Physics and Astronomy,\\
University of Rochester,\\
Rochester, NY 14627-0171\\
USA\\
\\
and\\
\\
Ziemowit Popowicz \\
Institute of Theoretical Physics, \\
University of Wroclaw,\\
50-205 Wroclaw\\ 
Poland.}
\date{}
\maketitle

\begin{abstract}

In this letter, we study systematically the general properties of the
$B$-extension of any integrable model. In addition to discussing the
general properties of Hamiltonians, Hamiltonian structures etc, we
also clarify the origin of \lq\lq exotic'' charges in such models. We
show that, in such models, there exist at least two sets of non-local
conserved charges (and more if $N>1$ supersymmetry is present) and
that the \lq\lq exotic'' charges are part of this non-local charge
hierarchy. The construction of these non-local charges from the Lax
operator is explained.

\end{abstract}

\vfill\eject
\section{Introduction:}

Integrable models \cite{IM} appear naturally in the study of strings in the
matrix model approach. Thus, while the KdV hierarchy is obtained in
the double scaling limit of the one matrix model \cite{DS}, the supersymmetric
matrix models lead to a particular supersymmetric version of the KdV
hierarchy known as the $N=1$ supersymmetric KdV-B hierarchy
\cite{BB}. In  simple
terms, if $u$ denotes the dynamical variable of the KdV equation
\[ u_{t} = u_{xxx} + 6 u u_{x}, \]
where the subscripts represent differentiation with respect to the
corresponding variables, then, the $N=1$ supersymmetric KdV-B hierarchy is
given  by
\[
 \Phi_{t}  =  \Phi_{xxx} + 3 (D(D\Phi)^{2}).
\]
Here, $\Phi(x,\theta) = \psi(x) + \theta u(x)$ represents the
dynamical variable which is a $N=1$ fermionic superfield with $\theta$
denoting the Grassmann coordinate and
\[ D = {\partial\over \partial\theta} + \theta {\partial\over \partial
x} .\]
There are, of course, other supersymmetrizations of the KdV hierarchy
that are integrable \cite{MRM}, but it is this particular
supersymmetrization \cite{FS}
that manifests in the study of string theories. Therefore, in this
letter, we undertake a systematic study of the properties of such a
supersymmetrization. In particular, we show, in section {\bf 2}, that
this  particular method of supersymmetrization can be applied to any
integrable model, although the original study involved the bosonic KdV
hierarchy. In section {\bf 3}, we bring out some general properties of
such models, such as the Hamiltonians, Hamiltonian structures,
recursion operators etc. In these models, there arise local conserved
charges which have opposite Grassmann parity relative to the
Hamiltonians of the system. The origin of such \lq\lq exotic'' charges
\cite{FS} is explained in
section {\bf 4}, where we identify that such local charges belong to
the hierarchy of  an infinite  set of non-local charges. In fact, we show that,
in such models, there exist, at least, two infinite sets of non-local
charges and may be more. Explicitly, in the $N=2$ supersymmetric KdV-B
hierarchy, we show that there exist three infinite sets of non-local
charges and present a method for constructing them. In section {\bf
5}, we present briefly an alternate description for such a
supersymmetrization which allows the construction of $B$-extensions of
systems such as the NLS and the AKNS hierarchies. A brief conclusion
is presented in section {\bf 6}. We used the symbolic computer language 
Reduce \cite{Red} and the special package \cite{Zp1} in all calculations
presented in this letter. 

\section{Model:}

Let us consider a general integrable model of the form
\begin{equation}
\phi_{t} = (A[\phi])_{x},
\end{equation}
where the subscripts refer to differentiation with respect to the
corresponding variables. Here, $\phi$ is a general dynamical variable.
It can be a purely bosonic function of $x$ alone, in which case, the
equation will represent the dynamics of a bosonic integrable
system ($N=0$ supersymmetry). Alternately, $\phi$ may represent a
superfield (bosonic or fermionic) depending on $x$ as well as $N$
fermionic coordinates, $\theta_{i}, i=1,2,\cdots ,N$, in
which case, the dynamical equation will describe  an integrable
model with $N$-extended supersymmetry. Let us denote the covariant
derivatives with respect to the $N$ fermionic coordinates by
\begin{equation}
D_{i} = {\partial\over \partial \theta_{i}} + \theta_{i}{\partial\over
\partial x},\qquad\quad i=1,2,\cdots ,N,
\end{equation}
which satisfy
\[ \left[D_{i},D_{j}\right]_{+} = \delta_{ij}\,{\partial\over \partial
x} = \delta_{ij}\,\partial. \]
We can now introduce a new superfield, ($\phi$ and
$\tilde{\phi}$ are superfields depending on the original $N$ fermionic
coordinates)
\begin{equation}
G=G(x,\theta_{1},\cdots ,\theta_{N+1}) = \tilde{\phi} +
\theta_{N+1}\phi,
\end{equation}
which depends on one extra fermionic coordinate and has opposite
Grassmann parity relative to $\phi$ (the original dynamical variable),
and define the dynamical equation
\begin{equation}
(D_{N+1}G)_{t} = (A[(D_{N+1}G)])_{x},
\end{equation}
which would represent an integrable system with $(N+1)$-extended
supersymmetry. This, therefore, describes the generalization of
Beckers' extension to (extended) supersymmetric models. (Basically,
the original $\phi$ equation remains unchanged under this extension
since $\left.(D_{N+1}G)\right|_{\theta_{N+1}=0}=\phi$.)

Thus, for example, with $\phi = u(x)$ and $A[u] = u_{xx}+3u^{2}$, we
have the bosonic KdV equation ($N=0$ supersymmetry) while, with $G =
\Phi(x,\theta)$, where $\Phi$ is a fermionic superfield, the equation
\begin{eqnarray}
(D\Phi)_{t} & = &
\left((D\Phi_{xx})+3(D\Phi)^{2}\right)_{x},\nonumber\\
{\rm or,{~~~}{~~~}}\quad \Phi_{t} & = & \Phi_{xxx} + 3D\left((D\Phi)^{2}\right),
\end{eqnarray}
represents the $N=1$ supersymmetric KdV-B equation \cite{BB}. Similarly, for
$\Phi(x,\theta_{1})$ a 
fermionic superfield and $A[\Phi] = -(\Phi_{xx}+3\Phi(D_{1}\Phi))$, the
dynamical equation represents the $N=1$ supersymmetric KdV equation \cite{MRM},
while, with $G(x,\theta_{1},\theta_{2})$ a bosonic superfield, the
equation
\begin{eqnarray}
(D_{2}G)_{t} & = &
-\left((D_{2}G_{xx})+3(D_{2}G)(D_{1}D_{2}G)\right)_{x},\nonumber\\
{\rm or,{~~~}{~~~}}\quad G_{t} & = & - G_{xxx} -
3D_{2}\left((D_{2}G)(D_{1}D_{2}G)\right),
\end{eqnarray}
would give rise to an $N=2$ extended supersymmetric KdV equation of
the $B$-type. Similarly, if $\phi$ represents an $N=2$ superfield and
the dynamical equation gives the $N=2$ supersymmetric KdV equation
\cite{LM,Zp2},
then, the $G$ equation would correspond to the $N=3$ supersymetric
KdV-B equation and so on. While this procedure is quite general, in
this letter, we would study the specific model in eq. (6) and bring out
properties of this model which are nonetheless common to all such
models. We also note here that, as described above, this extension,
when applied twice to a given equation, would seem to lead to a nonlocal
dynamical equation and, therefore, is not useful. Similarly, if the
right hand side of eq. (1) is not a total space derivative, this
method will also appear to fail.  We would come back
to this point in section {\bf 5}, where we would describe an alternate, but
equivalent generalization, which can be applied to any given equation and as
many times, without introducing non-locality. 

\section{Hamiltonians and Hamiltonian Structures:}

Let us next look at the general model of eq. (1). We note that if
\begin{equation}
H_{n}^{(N)} = \int dx\, d\theta_{1}\cdots
d\theta_{N}\,h_{n}^{(N)}[\phi],\qquad n=1,2,\cdots ,
\end{equation}
represent the Hamiltonians of the original model, then,
\begin{equation}
H_{n}^{(N+1)} = \int dx\, d\theta_{1}\cdots
d\theta_{N+1}\,h_{n}^{(N)}[(D_{N+1}G)],\qquad n= 1,2,\cdots ,
\end{equation}
would correspond to the Hamiltonians of the extended $B$-model
\cite{FS}. These
are conserved local quantities which would be invariant under the
extended  supersymmetry and we note that,
since the integration, in the second case, is over an additional
fermionic  variable relative to the definition of the original
charges, the Hamiltonians of the new system 
would have an opposite Grassmann parity compared to those of the
original system. It is also not hard to see that the Hamiltonian
densities can be written as
\begin{equation}
h_{n}^{(N)}[(D_{N+1}G)] = h_{n}^{(N)}[\phi] +
\theta_{N+1}\tilde{h}_{n}^{(N+1)}[\phi,\tilde{\phi}],
\end{equation}
so that each of the two parts of the Hamiltonians, namely, the
$\theta_{N+1}$ independent term  as well as the linear term in
$\theta_{N+1}$, will be independently conserved. However, the
$\theta_{N+1}$ independent term in the density would give a conserved
charge (when integrated over appropriate coordinates) which is
invariant only under the lower, $N$-extended supersymmetry.

The Hamiltonian structures of the two systems are also related in a
simple manner. Suppose ${\cal D}^{(N)}[\phi]$ represents the Hamiltonian
structure of the original system so that we can write
\begin{equation}
\phi_{t} = {\cal D}^{(N)}[\phi]\,{\delta H_{n}^{(N)}[\phi]\over \delta\phi}.
\end{equation}
Then, it follows that, we can write
\begin{eqnarray}
(D_{N+1}G)_{t} & = & {\cal D}^{(N)}[(D_{N+1}G)]\,{\delta H_{n}^{(N+1)}\over
\delta (D_{N+1}G)},\nonumber\\
{\rm or,}{~~~~~}\quad G_{t} & = & {\cal D}^{(N+1)}[G]\,{\delta
H_{n}^{(N+1)}\over\delta G},
\end{eqnarray}
where
\begin{equation}
{\cal D}^{(N+1)}[G] = D_{N+1}^{-1}\,{\cal D}[(D_{N+1}G)]\,D_{N+1}^{-1}.
\end{equation}
We note here that the new Hamiltonian structure would have an opposite
Grassmann parity from the old one, simply because of the delta
functions involving fermionic coordinates. (Such Hamiltonian
structures are known as anti-brackets or Buttin brackets
\cite{AB}.) This is  consistent with the
change of the Grassmann parity for the Hamiltonians that we have
already noted. The recursion operators for the two systems are
similarly related and, without going into details, we simply note here
that
\begin{equation}
R^{(N+1)}[G] = D_{N+1}^{-1}\,R^{(N)}[(D_{N+1}G)]\, D_{N+1}.
\end{equation}

The Lax description for the two systems are also simply related. For
example, we know that the Lax description for the KdV hierarchy is
given in terms of the Lax operator of the form
\[ L = \partial^{2} + u ,\]
where $\partial$ represents ${\partial\over \partial x}$. It follows,
then, that the Lax operator
\begin{equation}
L = \partial^{2} + (D\Phi),
\end{equation}
would describe the $N=1$ supersymmetric KdV-B hierarchy through the same normal
Lax representation,
\[ {\partial L\over \partial t} = \left[ L, (L^{3/2})_{+}\right] \].

Similarly, the $N=1$ supersymmetric KdV equation can be described
either by a standard representation with the Lax operator \cite{MRM}
\[ L = \partial^{2} + D_{1}\Phi, \]
or by a Lax operator \cite{BD1}
\[ L = \partial + D_{1}^{-1}\Phi, \]
with the non-standard Lax representation
\[ {\partial L\over \partial t} = \left[ L, (L^{3})_{\geq 1}\right]. \]
Correspondingly, the $N=2$ supersymmetric $B$-extension of this system
can also have a standard as well as a non-standard representation
through the Lax operators
\begin{eqnarray}
L^{\rm std} & = & \partial^{2} + D_{1} (D_{2}G),\nonumber\\
L^{\rm nstd} & = & \partial + D_{1}^{-1} (D_{2}G).
\end{eqnarray}

As opposed to these Lax operators it is also possible to define the Lax 
operator on the $N=2$ superspace. Indeed 
the method of supercomplexification \cite{Zp3} provides a much more general 
procedure for obtaining the
$B$-extensions and applied to this model, it provides the Lax
operator 
\[ L^{\rm sc} = \partial +D_{1}^{-1} (D_2G) - D_{1}^{-1}G_xD_2^{-1},\]
which would describe the system with a non-standard Lax representation.

The conserved Hamiltonians of the system can be obtained from the
super residues of any of these three Lax operators. Thus, for example,
the Hamiltonians of the system can be written in terms of the
non-standard Lax operator as
\begin{equation}
H_{n} = \int dx\,d\theta_{1}\,d\theta_{2}\; sRes (L^{\rm
nstd})^{2n-1} = \int dx\,d\theta_{1}\,d\theta_{2}\,h_{n},
\end{equation}
and the first two nontrivial charges of the series have the forms
\begin{eqnarray*}
H_{3} & = & \int dx\,d\theta_{1}\,d\theta_{2}\, G (D_{1}G_{x}),\\
H_{5} & = & \int dx\,d\theta_{1}\,d\theta_{2}\,\left[G (D_{1}G_{xxx}) +
4 G (D_{1}G_{x}) (D_{1}D_{2}G)\right].
\end{eqnarray*}

It is worth noting here that the $N=2$ supersymmetric $B$-extension
(eq. (6)) has yet another Lax representation \cite{DGS}. Namely, consider the
Lax operator 
\begin{equation}
L = D_{1} + \partial^{-1} D_{2}G - G D_{2}\partial^{-1}.
\end{equation}
Then, it is straight forward to check that the non-standard Lax
equation
\begin{equation}
{\partial L\over \partial t} = \left[L, (L^{6})_{\geq 1}\right],
\end{equation}
gives the $N=2$ equation of eq. (6). However, this Lax operator,
surprisingly, does not yield any of the conserved charges of the
system. This is indeed a puzzling feature which deserves further study. 

\section{Non-local Charges}

Let us now concentrate on the $N=2$ model of the previous section (eq.
(6)) for
concreteness, although the features we are going to discuss are quite
general. We note that although the Hamiltonians for this system are
fermionic, as we have discussed, there are also the following bosonic
charges which can be explicitly checked to be conserved, namely,
\begin{eqnarray}
\tilde{H}_{1} & = & \int dx\,d\theta_{1}\,d\theta_{2}\; G,\nonumber\\
\tilde{H}_{2} & = & \int dx\,d\theta\,d\theta_{2}\; G^{2},\nonumber\\
\tilde{H}_{3} & = & \int dx\,d\theta_{1}\,d\theta_{2}\; \left({1\over
3}G^{3} - G(D_{1}D_{2}G)\right).
\end{eqnarray}
At first sight, the existence of such charges with opposite Grassmann
parity would seem surprising and, in fact, the existence of such a
charge was already noted earlier in connection with the $N=1$
supersymmetric KdV-B system and was termed \lq\lq exotic'' \cite{FS}. In what
follows,  we would try to clarify
the origin of such charges and show that, in such systems, there are,
at least, two infinite sets of conserved non-local charges (and may be
more) and that these \lq\lq exotic'' charges are part of an infinite
hierarchy of non-local charges.

To begin with, let us recall that supersymmetric integrable systems,
in general, possess non-local charges \cite{K,DM,BD2,BD1}. However, in
the study  of
dispersionless limits of $B$-extended  models, it was
already noted \cite{BCD} that there are two sets of conserved,
non-local charges 
present. In fact, a little bit of analysis shows that the
$B$-extensions of integrable models will always
have at least two infinite sets of non-local conserved charges. For example, in
the $N=2$ supersymmetric KdV-B hierarchy, let us note that the charges
\begin{equation}
\overline{H}_{n} = \int
dx\,d\theta_{1}\,d\theta_{2}\;\left(D_{2}^{-1}\,sRes
(L^{\rm nstd})^{2n-1}\right) = \int
dx\,d\theta_{1}\,d\theta_{2}\,(D_{2}^{-1} h_{n}),
\end{equation}
will be conserved simply because these correspond to the conserved
charges of the original system and the original equations are
unmodified by this extension. Such a hierarchy of charges will always
be present. (In the spirit of eq. (9), these charges would be obtained
from the $\theta_{2}$ independent part of the densities.) They are
manifestly  non-local and, consequently, are not
invariant under the $N=2$ extended supersymmetry. Let us also note
that, by definition,
\[\int dx\,d\theta_{1}\,d\theta_{2}\,\left(\partial^{-1}D_{1}D_{2}
sRes (L^{nstd})^{2n-1}\right) = 0. \]
There is also a second set of non-local charges which one can
construct. Namely, let us evaluate the square root of the Lax operator
for the non-standard representation. Conventionally, the super
residues of the odd powers of the square root of the Lax operator
gives rise to conserved non-local charges in supersymmetric integrable
models. As we will now show, the system under study presents a novel
feature and, consequently, leads to new charges. Let us note that
since both $D_{1}$ and $D_{2}$
satisfy
\[ D_{1}^{2} = \partial = D_{2}^{2} ,\]
it follows that the general form of the square root can be determined
to have the form
\begin{eqnarray}
(L^{\rm nstd})^{1/2} & = & \alpha D_{1} + \beta D_{2} + 2\alpha
(D_{2}^{-1}G) - (\alpha (\partial^{-1} D_{1}D_{2} G) + \beta
G)D_{1}^{-1}\nonumber\\
 &  &  + \left(\alpha (D_{2}G) - \beta (D_{1}G) - \beta
(D_{2}^{-1}G^{2})\right)\partial^{-1}\nonumber\\
 &  &  +{1\over 2} \left(\alpha
(\partial^{-1}D_{1}D_{2}G)^{2} + \beta
(\partial^{-1}D_{1}D_{2}G^{2})\right)D_{1}^{-3} + \cdots ,
\end{eqnarray}
where the constant parameters $\alpha$ and $\beta$ are constrained to
satisfy $\alpha^{2}+\beta^{2}=1$. While non-local charges have been
constructed earlier from square (and quartic) roots \cite{DM,BD1,BD2},
here  we have the
novel feature that there is a one parameter family of square roots of the
Lax operator. We think this feature would exist in
extended supersymmetric models with $N>1$. In fact, let us note that
for $\alpha = 1$ and $\beta = 0$, this square root coincides with what
has been calculated earlier \cite{BD1,BD2}. But, this is, in fact, the
more  general form of the square root with a richer structure.

From the structure of this general square root, let us note that we
can construct conserved charges by taking the \lq\lq $sRes$'' of odd
powers of the square root and would, in general, give non-local
charges. In fact, let us note that the first few of these charges have
the forms ($dz = dx\,d\theta_{1}\,d\theta_{2}$)
\begin{eqnarray}
\int dz\, sRes (L^{\rm nstd})^{1/2} & = & -\beta \int
dz\,G, \nonumber\\
\int dz\, sRes (L^{\rm nstd})^{3/2} & = & - {3\alpha\over 2} \int
dz\, G^{2},\nonumber\\
\int dz\, sRes (L^{\rm nstd})^{5/2} & = & \int dz\left[2\beta\left({1\over 3}
G^{3} - G (D_{1}D_{2}G)\right) + \beta
\left(D_{2}^{-1}(G_{x}(D_{1}G))\right)\right.\nonumber\\
 & & - \left.{\alpha\over 2}\left({1\over 3}
(D_{1}^{-1}D_{2}G)^{3} + 2
(D_{1}^{-1}((D_{2}G)(D_{1}D_{2}G)))\right)\right].
\end{eqnarray}
Thus, we see that the \lq\lq $sRes$'' of the odd powers of the square
root leads to conserved charges which are a combination of new
non-local charges (the first few of which are local) as well as  old
ones of  the form eq. (20). In fact, if we
neglect the old non-local charges in these expressions, we see that
the one parameter family of charges really leads to two distinct sets
of conserved charges. Thus, for example, when $\alpha =1$ (and,
therefore, $\beta = 0$), the non-local charges coincide with what has
been obtained earlier \cite{DM,BD1,BD2}. However, when $\beta = 1$, we
have a  new set of
non-local conserved charges for the system. Thus, we conclude that, in
this $N=2$ supersymmetric model, we have, in fact, three sets of
conserved non-local charges. Furthermore, we now recognize that the
three \lq\lq exotic'' charges belong to this hierarchy of non-local
charges and can only be obtained if we take the general square
root. (In other words, the first few members of the non-local
hierarchy of charges is really local, even though higher order ones
are truly non-local. We have explicitly verified with REDUCE that
there are no more local \lq\lq exotic'' (bosonic) conserved charges
present.)

To complete the story of the \lq\lq exotic'' charges, let us look at
the simpler system of $N=1$ supersymmetric KdV-B hierarchy. Here the Lax
operator, as we have seen in eq. (14), has the form
\[ L = \partial^{2} + (D\Phi). \]
The fermionic Hamiltonians of the system are given by
\begin{equation}
H_{n} = \int dx\,d\theta\, Res (L^{(2n-1)/2}) = \int
dx\,d\theta\,h_{n},\qquad\qquad n=1,2,\cdots, 
\end{equation}
and the first set of non-local charges are given by
\begin{equation}
\overline{H}_{n} = \int dx\,d\theta\,(D^{-1}Res (L^{(2n-1)/2})) = \int
dx\,d\theta\,(D^{-1}h_{n}).
\end{equation}
Since, in this case, we have only one fermionic coordinate, the
quartic root is without any arbitrary parameter and the residues of the
odd powers of it give rise to a linear combination of new non-local
conserved charges and the ones in eq. (24). Ignoring these old charges, we
can write the new set of non-local charges to be coming from
\begin{equation}
\tilde{H}_{n} = \int dx\,d\theta\, Res (L^{(2n-1)/4}),
\end{equation}
These are bosonic charges and explicitly, the first few of them have
the form
\begin{eqnarray}
\tilde{H}_{1} & = & {1\over 2}\int dx\,d\theta\, \Phi,\nonumber\\
\tilde{H}_{2} & = & {1\over 4}\int dx\,d\theta\, (D^{2}\Phi) = 0,\nonumber\\
\tilde{H}_{3} & = & {1\over 4} \int dx\,d\theta\, \Phi
(D\Phi),\nonumber\\
\tilde{H}_{4} & = & {3\over 8} \int dx\,d\theta\, \Phi
(D^{3}\Phi),\nonumber\\
\tilde{H}_{5} & = & {1\over 8} \int dx\,d\theta\,
\Phi\left((D^{5}\Phi) + 2 (D\Phi)^{2}\right).
\end{eqnarray}
Of these, only $\tilde{H}_{3}$ was found earlier and termed \lq\lq
exotic'' \cite{FS}. We see that it belongs to a hierarchy of non-local
charges, the first four of which are, in fact, local. (We suspect
that, in this particular case, this new set of charges is indeed
local. This follows from an analysis of the structure of the charges
in the dispersionless limit \cite{BCD}. However, this is not a general
feature.)

\section{Alternate description:}

As we had noted earlier, the conventional $B$-extension cannot be
applied to equations where the time evolution of the dynamical
variable is not a space derivative. Furthermore, even when the
$B$-extension exists, it gives
non-local equations if applied more than once. In this section, we
will describe very briefly, an alternate  extension
which does not suffer from this problem. Let us consider an integrable
system of the form 
\begin{equation}
\phi_{t} = B[\phi].
\end{equation}
If we now define a superfield
\begin{equation}
\overline{G} (x, \theta_{1},\cdots, \theta_{N+1}) = \phi +
\theta_{N+1}\overline{\phi},
\end{equation}
then, we note that the new superfield depends on one extra fermionic
coordinate and has the same Grassmann parity as the original variable
$\phi$. If we now define a dynamical system described by
\begin{equation}
\overline{G}_{t} = B [\overline{G}]. 
\end{equation}
then, this system would be integrable. We note that the $\theta_{N+1}$
independent part of this equation would correspond to eq. (27), the
original equation, so that
it provides an alternate description of the $B$-extension. In fact,
when we can write $B[\phi] = (A[\phi])_{x}$,
the two descriptions would be equivalent and will map into each other under
\begin{equation}
\overline{G} \rightarrow (D_{N+1}G).
\end{equation}
All the discussions of the earlier sections can be carried out in this
framework as well. However, the advantage of such a description may
lie in the fact that the $B$-extended equation, in eq. (29), is
exactly  like the original equation independent of whether the right
hand side is a total derivative or not. Consequently, we can apply
$B$-extension  to any given equation and as many times without running
into the problems
of non-locality. (In simple terms, the new variables in the alternate
representation are more local than the older ones.) As a result,
systems, such as the NLS and the AKNS hierarchies, which were thought not to
have a local $B$-extension (in the standard approach) \cite{BD3,AD},
can actually have one in this alternate description.

\section{Conclusion}

In this letter, we have studied systematically the general features of
$B$-extension of any given integrable system. We have brought out general
features such as the Hamiltonians, Hamiltonian structures and recursion
operators. We have clarified the origin of \lq\lq exotic'' charges in
such models and have identified them as belonging to an infinite set
of non-local charges. We have shown that, in such models, there
naturally exist, at least, two infinite sets of non-local charges and,
for $N>1$ supersymmetry, even more. We have explicitly shown that the
$N=2$ supersymmetric KdV-B hierarchy has three sets of non-local conserved
quantities and have discussed their construction starting from the Lax
operator.

\section*{Acknowledgments}

A.D. acknowledges support in part by the U.S. Dept. of
Energy Grant  DE-FG 02-91ER40685 while Z.P. is supported in part by
the  Polish KBN  Grant  2 P0 3B 136 16.


\begin{thebibliography}{99}
\bibitem{IM} G.B. Whitham, {\it Linear and Nonlinear Waves}, John
Wiley, 1974; L.D. Faddeev and L.A. Takhtajan, {\it Hamiltonian Methods
and the Theory of Solitons}, Springer, 1987; A. Das, {\it Integrable
Models}, World Scientific, 1989.
\bibitem{DS} D.J. Gross and A.A. Migdal, Phys. Rev. Lett. 64 (1990)
127; D.J. Gross and A.A. Migdal, Nucl. Phys. B340 (1990)333;
E. Br\'{e}zin and V.A. Kazakov, Phys. Lett. B236 (1990) 144;
M. Douglass and S.H. Shenker, Nuc. Phys. B335 (1990) 635.
\bibitem{BB} K. Becker and M. Becker, Mod. Phys. Lett. A8 (1993) 1205.
\bibitem{MRM} Yu. I. Manin and A. O. Radul, Comm. Math. Phys. {\bf 98}
(1985) 65; P. Mathieu, Phys. Lett. {\bf B203} (1988) 287;
P. Mathieu, J. Math. Phys. {\bf 29} (1988) 2499.
\bibitem{FS} J.M. Figueroa-O'Farrill and S. Stanciu, Phys. Lett. B316
(1993) 282.
\bibitem{Red} A. Hearn Reduce {\it User's Manual 3.7}, (1999).
\bibitem{Zp1} Z. Popowicz Compt. Phys. Commun {\bf 100} (1997) 277. 
\bibitem{LM} P. Laberge and P. Mathieu Phys. Lett {\bf B215} (1988) 718; 
P. Labelle and P. Mathieu J. Math. Phys. {\bf 32} (1991) 923.   
\bibitem{Zp2} Z. Popowicz  J. Phys. {\bf A29} (1996) 4987; Z. Popowicz
Phys. Lett {\bf A174} (1993) 411.
\bibitem{AB} D.A. Leites, Dokl. Akad. Nauka SSSR {\bf 236} (1977) 804
(in Russian); A. Batalin and G.A. Vilkovsky, Phys. Lett. {\bf B102}
(1981) 27; A. Batalin and G.A. Vilkovsky, Nuc. Phys. {\bf B234} (1984) 106.
\bibitem{BD1} J.C. Brunelli and A. Das, Phys. Lett. {\bf B337} (1994)
303; J.C. Brunelli and A. Das, Int. J. Mod. Phys. {\bf A10} (1995)
4563.
\bibitem{Zp3} Z. Popowicz  Phys. Lett. {\bf B459} (1999) 150.
\bibitem{DGS} F. Delduc, L. Gallot and A. Sorin Nucl. Phys {\bf B558}
(1999) 535. 
\bibitem{K} P.H.M. Kersten, Phys. Lett. {\bf A134} (1988) 25.
\bibitem{DM} P. Dargis and P. Mathieu, Phys. Lett. {\bf A176} (1993)
67.
\bibitem{BD2} J.C. Brunelli and A. Das, Phys. Lett. {\bf B354} (1995) 307.
\bibitem{BCD} J. Barcelos-Neto, A. Constandache and A. Das,
Phys. Lett. {\bf A268} (2000) 342.
\bibitem{BD3} J.C. Brunelli and A. Das, Phys. Lett. {\bf B409} (1997)
229.
\bibitem{AD} H.A. Aratyn and A. Das, Mod. Phys. Lett. {\bf A15} (1998) 1185.

\end{thebibliography}
\end{document}